\pdfoutput=1
\documentclass{PoS}

\usepackage{amsmath}
\usepackage{amssymb}
\usepackage{euscript}
\usepackage{bbm}
\usepackage{bm}
\usepackage[T1]{fontenc}
\usepackage[utf8]{inputenc}

\usepackage[pdftex]{graphicx}
\usepackage{mathptmx}
\usepackage{times}

\title{The Exclusive NLO DGLAP Kernels for Non-Singlet Evolution
\thanks{  This work is partly supported by the EU 
  Framework Programme grants MRTN-CT-2006-035505 and
  MTKD-CT-2004-014319
  and by the Polish Ministry of Science and Higher Education grant 
  No.\ 153/6.PR UE/2007/7}}

\ShortTitle{Exclusive NLO DGLAP kernels }

\author{\speaker{Maciej Skrzypek}\\
 H. Niewodniczański Institute of Nuclear Physics, Polish
  Academy of Sciences,\\
  ul.\ Radzikowskiego 152, 31-342 Cracow, Poland,\\
        E-mail: \email{Maciej.Skrzypek@ifj.edu.pl}}

\author{Stanislaw Jadach \\
 H. Niewodniczański Institute of Nuclear Physics, Polish
  Academy of Sciences,\\
  ul.\ Radzikowskiego 152, 31-342 Cracow, Poland,\\
        E-mail: \email{Stanislaw.Jadach@ifj.edu.pl}}

\abstract{
We show for a first time ever a prototype of a fully exclusive QCD NLO
 parton shower for the initial state (albeit for a limited set of
 diagrams). It is based on the rigorous
 theorems of the collinear factorisation, however the standard DGLAP
 evolution kernels are replaced by their exclusive versions. Contrary
 to the standard DGLAP approach, the constructed parton shower
 provides fully exclusive events, i.e.\ four-momenta of all
 emitted partons. At the inclusive level it is identical to the
 $\overline{MS}$ scheme of the DGLAP evolution.

\begin{flushleft}
\vskip 1cm
\normalsize \hskip -0.7cm
\bf IFJPAN-IV-2009-7\\
\end{flushleft}
}

\FullConference{The 2009 Europhysics Conference on High Energy Physics,\\
		 July 16 - 22 2009\\
		 Krakow, Poland}

\begin{document}

With the approach of the first data from the LHC experiments the
issue of the precision of the Monte Carlo (MC) simulations of the
perturbative QCD becomes more and more important. In the domain of the
semi-inclusive, analytical or semi-analytical calculations the results
are spectacular: the NLO accuracy is a standard and numerous
quantities are calculated to much higher precision, just to mention
the DGLAP kernels known to NNLO level \cite{Moch:2004pa}. 
On the contrary, all the currently 
available MC Parton Shower (PS) codes are of the
LO or, at most, of the improved LO type (as far as the differential
distributions are concerned, the overall, inclusive, normalization can
easily be corrected to higher order precision), based on the methods
developed in early 1980-s, cf.\ for
example \cite{Frixione:2006he,Sjostrand:2007gs,Bahr:2008pv}.

It is therefore justified to ask whether it is possible to develop a
new, more precise scheme for the QCD MC parton shower? In this
presentation we will demonstrate that yes, it is feasible. We will
describe a working prototype of such a genuine NLO parton shower for
the initial state QCD which we have developed recently within the {\tt
KRKmc} project. %
Note that there are some other attempts to
construct parton shower schemes beyond
leading order as well,
cf.\ eg.\ \cite{Deak:2009ae,Collins:2007ph,Nagy:2007ty}.

The {\tt KRKmc} parton shower fulfills the following list of requirements:
\\
(A)
it is based firmly on Feynman Diagrams (Matrix Element) and
Lorentz-invariant phase space (LIPS);
\\
(B)
it is based rigorously on the collinear factorization 
\cite{Ellis:1978sf, Collins:1984kg};
\\
(C)
it implements {\em exactly} the NLO $\overline{MS}$ DGLAP
evolution \cite{Curci:1980uw}; 
\\
(D)
it implements fully unintegrated parton density functions (PDFs);
with NLO evolution done by MC itself,
using new, exclusive NLO kernels.
\\
The combination of the collinear factorization (item 1), which is the best
proven factorization scheme, with the complete kinematical information
of the generated partons (item 4), i.e.\ full exclusiveness of the
shower is the essence of the novelty of our approach.
Let us now highlight some of the important points in this new {\tt
KRKmc} scheme. For more details we refer to
Refs.\  \cite{Jadach:2009gm,Slawinska:2009gn,Jadach:2005bf}.%

{\bf Step 1.} 
Re-do the calculation of the NLO kernels, based on the framework of
Curci, Furmanski and Petronzio \cite{Curci:1980uw} but in the exclusive way. 
Use various types of evolution time
(virtuality, $k_\perp$, rapidity \dots ). The standard DGLAP kernels
do not depend on the choice of the evolution time. On the contrary,
the exclusive ones depend.
Use $\overline{MS}$ factorization scheme. 
In the Fig.\ 1 we show all the contributions to the
non-singlet LO and NLO
evolution kernels. In this work we restrict ourselves to the $C_F^2$
part (two ``bremsstrahlung-like'' graphs) labeled with the 
blue box. It is in a sense the most complicated part because it
involves the subtraction of the soft counter-term (labeled as $1-P$ in the
plot).
\begin{figure}
\label{fig:1}
\begin{center}
\includegraphics[width=15cm,height=8cm]{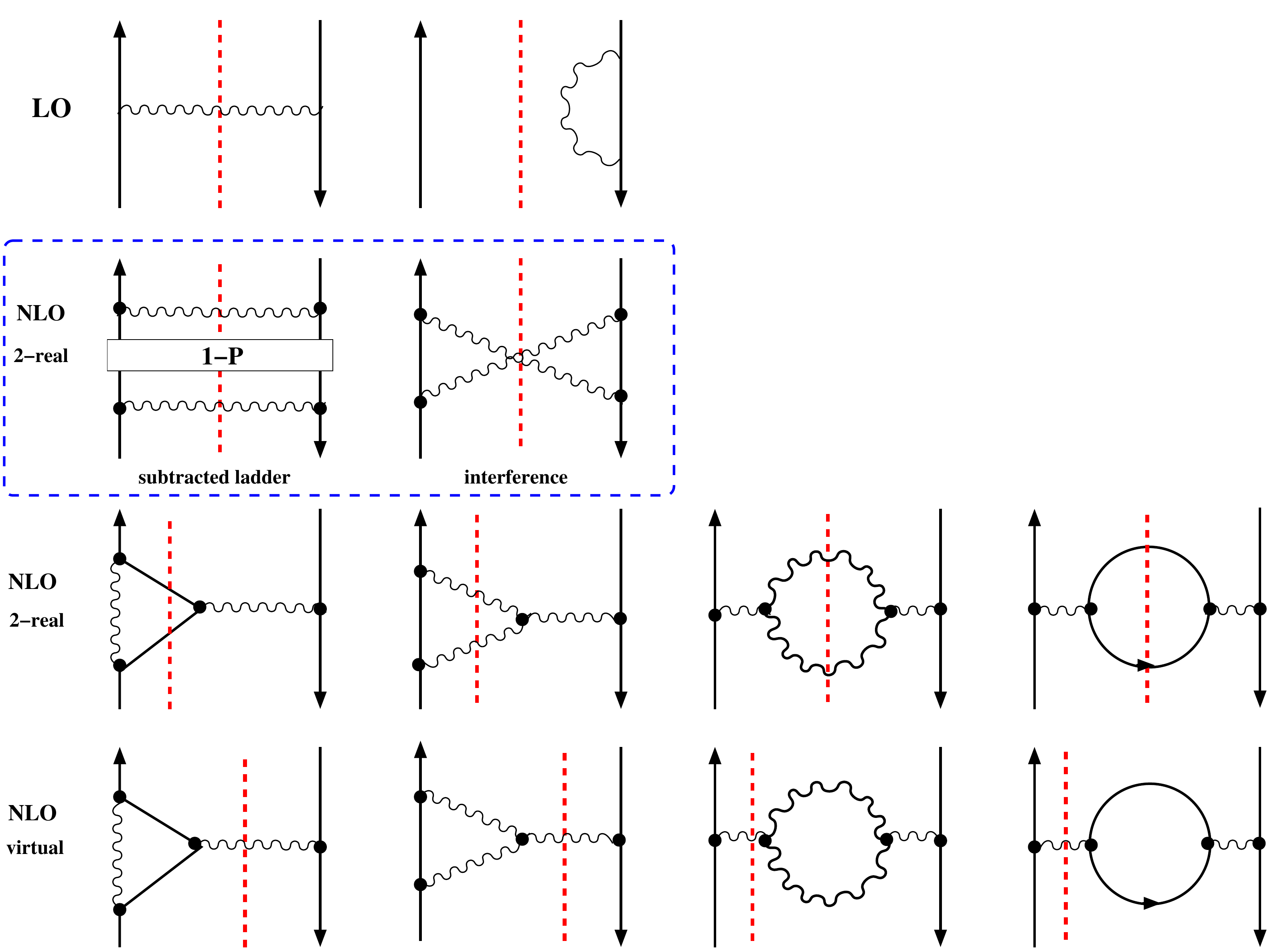}
\end{center}
\caption{Contributions to the NLO non-singlet evolution
kernel. The $C_F^2$ part label-led with the blue box.
        }
\end{figure}

{\bf Step 2.} Remove the dimensional regularization, go back to four-dimensions.
Use instead a geometrical cut-off {$\Delta$} for the collinear
singularity. But, all the time stay within 
the  $\overline{MS}$ scheme. This is possible because the collinear
pole comes from a simple, factorisable integral of the form
\begin{displaymath}
 \frac{1}{\epsilon}\Rightarrow
 \int_0^{Q^2} d\Bigl(\frac{q^2}{Q^2}\Bigr)\Bigl(\frac{Q^2}{q^2}\Bigr)^{1-\epsilon}
 \Rightarrow
 \int^{Q^2}_{\Delta^2} d\Bigl(\frac{q^2}{Q^2}\Bigr) \frac{Q^2}{q^2}.
\end{displaymath} 

{\bf Step 3.}  Re-formulate the factorization formula, because the 
DGLAP equation mixes orders of perturbative expansion. The NLO DGLAP
kernel, denoted as $P$, is in fact a sum of LO and NLO
pieces. Therefore $P^k$ terms are a mixture of various orders:
\begin{displaymath}
P=\alpha P^{LO} + \alpha^2 P^{NLO}
\;\;
\Rightarrow\;\;
P^2 = \alpha^2( P^{LO})^2 
  +\alpha^3 ( P^{LO}P^{NLO}+ P^{NLO}P^{LO})
  +\alpha^4 ( P^{NLO})^2.
\end{displaymath}
Such a set-up, with partial, incomplete higher orders, may lead for
example to the negative events in the MC and should be avoided.

{\bf Step 4.} Resign from the ordering in the evolution time in the
underlying LO crude MC. Use the Bose-symmetric form instead:
\begin{align}
\int_{t_{\min}}^{t_{\max}}\prod_i^N dt_i\, \theta_{t_i>t_{i-1}}
\;\;
\Rightarrow\;\;
\frac{1}{N!}\int_{t_{\min}}^{t_{\max}}\prod_i^N dt_i. 
\end{align}
The ordering is an approximate feature of the LO matrix element and
therefore it is not strict at the NLO level and we have to explicitly sum
over entire phase space.

{\bf Step 5.}
Construct appropriate MC weight with the NLO exclusive kernel. It will
be then applied on the top of the standard LO MC.

{\bf Step 6.}
Resolve the mismatch of the lower limits of the ``internal NLO phase
space'', i.e.\ of the internal degree of freedom integrated out in the
construction of the inclusive kernel. In the analytical calculation of
the $\overline{MS}$ kernel the lower limit of $dt_{internal}$
integration is set to 0, whereas in MC it is limited by some
$t_{\min}$ for all partons. In this work we resolve this conflict by
``artificial'' lowering of the $t_{\min}$ limit, below the actual
start of the evolution and by performing a LO pre-evolution in this
extended region. This way we maintain exact agreement with
the $\overline{MS}$ result.

We have implemented the above scheme of the exclusive NLO evolution in
the MC program {\tt KRKmc}. In the Fig.\ 2 we compare
its results with a standard NLO DGLAP MC evolution. Both evolutions
are implemented as weights on the top of the same LO Markovian MC
algorithm. The curves shown correspond to the contributions with one
and two NLO ``insertions''.
Evolution ranges from 10 GeV to 1 TeV.
LO pre-evolution starts at 1GeV from $\delta(1-x)$ distribution.
\begin{figure}
\centerline{
   \includegraphics[width=110mm,height=80mm]
                   {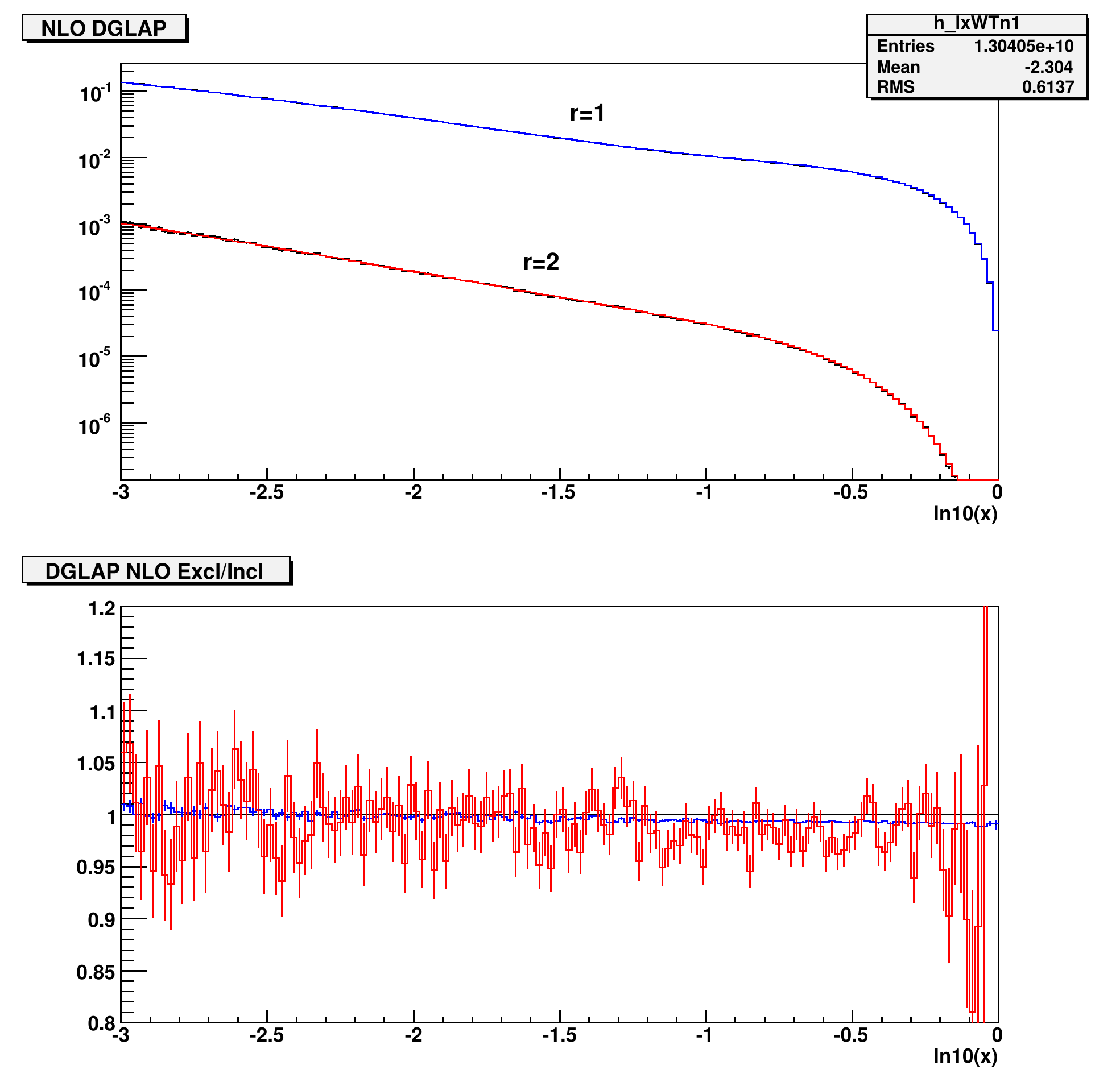}}
\caption{Comparison of exclusive and standard DGLAP NLO evolutions 
        }
\label{fig:porown}
\end{figure}
As one can see the agreement is very good, within the statistical
errors. This way we demonstrated for the first time ever that the QCD
NLO parton shower can be constructed (although for a limited set of
Feynman diagrams)!

We are currently in the process of adding the rest of graphs from
Fig.\ 1, omitted in this work. This way the non-singlet
evolution will be completed. Once also the singlet evolution is added,
we will be ready to construct the complete event generator for the
Drell-Yan-type processes at LHC or DIS at HERA.
\providecommand{\href}[2]{#2}
\end{document}